\documentclass[12pt,preprint]{aastex}

\newcommand{\etal}{et al.\ }

\shorttitle{X-rays from a brown dwarf TWA 5B}
\shortauthors{Y. Tsuboi et al.}

\received{2002 December 20}
\begin{document}

\title{Coronal X-ray emission from an intermediate-age brown dwarf}

\author{Yohko Tsuboi}
\affil{Department of Science and Engineering, Chuo University, 
Kasuga 1-13-27, Bunkyo-ku, Tokyo 112-8551, Japan}
\email{tsuboi@phys.chuo-u.ac.jp}

\author{Yoshitomo Maeda}
\affil{High Energy Astrophysics Division, Institute of Space and Astronautical Science, Yoshinodai 3-1-1, Sagamihara, Kanagawa 229-8510, Japan}
\email{ymaeda@astro.isas.ac.jp}

\author{Eric D. Feigelson, Gordon P. Garmire, George Chartas, Koji Mori}
\affil{Department of Astronomy and Astrophysics, The Pennsylvania State University, 525 Davey Laboratory, University Park, PA 16802, USA}
\email{edf@astro.psu.edu, garmire@astro.psu.edu, chartas@astro.psu.edu, mori@astro.psu.edu}

\author{Steven H. Pravdo}
\affil{Jet Propulsion Laboratory, Mail Stop 306-438, California Institute of Technology, Pasadena, CA 91109, USA}
\email{spravdo@jpl.nasa.gov}

\begin{abstract}

We report the X-ray detection of the brown dwarf (BD) companion TWA 5B
in a $\simeq 12$ Myr old pre-main sequence binary system. We clearly
resolve the faint companion (35 photons) separated from the X-ray
luminous primary by 2 arcsec in a {\it Chandra} ACIS image. TWA 5B
shows a soft X-ray spectrum with a low plasma temperature of only 0.3
keV and a constant flux during the 3 hour observation, of which the
characteristics are commonly seen in the solar corona. The X-ray
luminosity is 4$\times10^{27}$ erg s$^{-1}$ (0.1--10 keV band) or
$\log L_X/L_{bol} = -3.4$.

Comparing these properties to both younger and older BDs, we discuss
the evolution of the X-ray emission in BDs.  During their first few
Myr, they exhibit high levels of X-ray activity as seen in higher mass
pre-main sequence stars. The level in TWA 5B is still high at $t
\simeq 12$ Myr in $\log L_X/L_{bol}$ while $kT$ has already
substantially cooled.

\end{abstract}


\section{Introduction}

With masses below the hydrogen burning mass limit $M <$
0.08~M$_{\odot}$ = 80~M$_{jup}$ and cool, nearly neutral surfaces,
brown dwarfs (hereafter BDs) had been expected to lack high energy
activity. However, X-rays from several young BDs with ages $t \sim$1
Myr old were discovered with $ROSAT$ in nearby star forming regions
(Neuh\"auser \& Comer\'on 1998; Neuh\"auser, Briceno \&
Comer\'on. 1999, Mokler \& Stelzer 2002). After the launch of the
{\it Chandra X-ray Observatory}, the number of X-ray-detected young
BDs has increased an order of magnitude (Garmire et al.\ 2000,
Imanishi, Koyama \& Tsuboi 2001, Imanishi, Tsujimoto \& Koyama 2001,
Preibisch \& Zinnecker 2001, Feigelson et al.\ 2002a).
In young stellar clusters around 1 Myr old, hard and variable X-rays
are now detected from about 1/4 of the BD population near the
`saturation level' of $\log L_X/L_{bol} \simeq -3$.  These properties
are indistinguishable from those of higher mass young objects which
will evolve into dM stars.

Nearly all field L- and T-type older BDs with ages $t \simeq$
500--1000 Myr lack H$\alpha$ emission, a classical signature of active
regions and plage, but exhibit radio flares (Gizis et al.\ 2000,
Berger 2002). The weak H$\alpha$ in field BDs is particularly
surprising in light of their rapid rotation (Basri\ 2001). One field
BD at age $t \simeq 500$ Myr, however, is clearly magnetically active.
LP944-20 showed a 1--2 hr flare during a $Chandra$ observation with a
peak luminosity $L_X = 1 \times 10^{26}$ erg s$^{-1}$ ($\log
L_X/L_{bol} \simeq -4.1$) with a surprisingly soft spectrum, while it
exhibited no detectable quiescent X-rays before the flaring (Rutledge
et al.\ 2000). This BD also has exceptionally strong radio emission
(Berger et al.\ 2001) and shows H$\alpha$ emission of EW $\sim$ 1
{\AA} (Tinney 1998).  At the intermediate age of $t \simeq 100$ Myr,
three probable BDs in the Pleiades have been observed with
$XMM-Newton$ of which one, Roque 9, may have been detected at $\log
L_X/L_{bol} \simeq -2.6$ (Briggs \& Pye 2002, Stelzer \& Neuh\"auser
2002).

The evolution of magnetic activity in BDs thus appears confusing:
H$\alpha$ and X-rays appear to decay as they descend the Hayashi
tracks and cool, while radio emission does not. The nonthermal radio
emission is particularly puzzling with both flare and persistent
components far stronger than expected from the well-established
G\"udel-Benz radio/X-ray relationship that holds for all other
late-type stars (Berger 2002). This contrasts with the theoretical
expectation that BD surface magnetic activity should decay with age
due to increased resistivity of the atmosphere (Mohanty et al.\ 2002).
The evolution of magnetic activity in higher mass late-type stars is
also not fully understood. Various multiwavelength manifestations of
magnetic activity (X-ray flares, radio flares, H$\alpha$ emission,
large starspots, strong surface fields, etc.) are consistently seen at
high levels in T Tauri stars, but do not appear to depend on rotation
as seen in main sequence stars (Feigelson \& Montmerle 1999, Feigelson
et al.\ 2003).

We introduce to this issue the $Chandra$ detection of TWA 5B, a BD at
the intermediate age of $t \simeq 12$ Myr. TWA 5B is a faint red
object ($H = 12$ mag) located 2$''$ north of and co-moving with TWA 5A
which itself is a spectroscopic binary (Lowrance et al.\ 1999, Webb et
al.\ 1999, Neuh\"auser et al.\ 2000). This system is part of the
dispersed TW Hydra Association (TWA) of T Tauri stars, all of which
are likely outliers of the giant Sco-Cen OB association (Kastner et
al.\ 1997, Mamajek et al.\ 2000). TWA 5 is the first identified PMS
star-BD binary system with an estimated age of $12 \pm 6$ Myr
(Weintraub et al.\ 2000). We assume here the mean Hipparcos distance
to the 4 brightest TWA members of 55 pc. Optical and near-infrared
photometry of TWA 5B gives a spectral type of M8.5$-$M9 and bolometric
luminosity $\log L_{bol}/L_\odot \simeq -2.6 \pm 0.3$, giving an
inferred mass between 15 and 40 M$_{jup}$ (Neuh\"auser et al.\
2000). The equivalent width of H$\alpha$ $\sim$20{\AA} lies between
those of young BDs (10--100 {\AA} at $\sim$1 Myr old) and those of
field BDs ($\lesssim$ 1 {\AA} at $\sim$500--5000 Myr). TWA 5B is thus
clearly an intermediate-age BD and may be near the end of its brief
deuterium burning phase (Weintraub et al.\ 2000). The inferred mass is
not far from the boundary between a brown dwarf and a planet of about
12 M$_{jup}$.

\section{Observations and results}

TWA 5B was observed on 2001 April 15 with the {\it Chandra X-ray
Observatory} for 10.3 ks using the ACIS-S detector (see Weisskopf et
al.\ 2002 for a description of the satellite and instrument). The S3
CCD chip was operated in a rapid readout subarray mode to reduce
photon pileup of the primary TWA 5A which was known from $ROSAT$
studies to be very bright (Webb et al.\ 1999). The more frequent
readouts resulted in a net on-target exposure of 9.3 ks. Image
analysis started with Level 1 events from which we removed a $\pm$
0\arcsec.25 positional randomization, filtered with Good Time
Intervals, and corrected charge transfer inefficiency problem as
described by Townsley et al.\ (2000). To achieve the highest possible
spatial resolution, we applied a correction to estimate the arrival
location of each photon within each 0\arcsec.5 pixel based on the
precise distribution of charge received in adjacent pixels (Tsunemi et
al.\ 2001, Mori et al.\ 2001). Spectral analysis was performed using
Level 2 events provided by the standard satellite pipeline processing.

Visual examination of the raw (Level 2) ACIS S3 CCD image shows very
strong emission with about 6200 counts from TWA 5A, a CCD readout
stripe oriented NE/SW, and a faint excess of events 2\arcsec to the
north where TWA 5B lies. The TWA 5B emission becomes much clearer
after the special processing steps outlined above. Figure 1 (center)
shows this X-ray image in a very soft 0.1--1.2 keV band that
highlights the emission from the BD companion. Note that a resolution
of $\simeq$ 0\arcsec.3 FWHM is achieved using the subpixel process,
while the original data has $\simeq$ 0\arcsec.5 FWHM (these FWHM
values are measured from radial profile binned by 0.05 pixel.)

We extracted 35 photons from standard Level 2 event list around 
TWA 5B in the 0.1--8 keV band using a
0\arcsec.7 radius circle after removal of background from an annulus
around TWA 5A (see Figure 1, right panel). Since the extracted region
is smaller than the standard $Chandra$ point spread function at this
position, a modified `auxiliary reference file' was created. The
background-subtracted spectrum of TWA 5B was fitted with an optically
thin MEKAL thermal plasma model with elemental abundances fixed at 0.3
times of the solar value, based on the fitting results of other young
stellar objects (YSOs) detected by $Chandra$ and $ASCA$
(e.g. Imanishi, Koyama, \& Tsuboi 2001). The resulting spectrum, shown
in Figure 2, is unusually soft with peak emission at 0.7 keV and no
photons above 1.5 keV. Best-fit parameters with 90\% confidence levels
are: temperature $kT = 0.3^{+0.2}_{-0.1}$ keV, line-of-sight column
density $N_H < 6 \times 10^{20}$ cm$^{-2}$, emission measure $4 \pm 1
\times 10^{50}$ cm$^{-3}$, and luminosity $L_X = 4 \times 10^{27}$ erg
s$^{-1}$ in the 0.1--8 keV band. The spectral fit was essentially
perfect with reduced chi-squared of 0.5 for 3 degrees of freedom. The
plasma temperature and the luminosity are close to typical plasma
conditions in the Sun today: at solar maximum, typical temperatures
are $kT \simeq$ 0.2--0.5 keV with $L_X \simeq 1 \times 10^{27}$ erg
s$^{-1}$ (Peres et al.\ 2000). During a powerful solar flare, the
typical temperatures rise to $T \simeq$ 1--2 keV, much hotter than
seen in TWA 5B but typical of other PMS stars and BDs (e.g. Feigelson
et al.\ 2002a).

A light curve of the photon arrival times was examined and no
significant variations were found during the 11 ks duration of the
observation. We can exclude variations greater than 50\% in amplitude
over the 10 ks observation.

\section{Discussion}

\subsection{Evolution of X-ray emission }

As outlined in the Introduction, there is considerable confusion in
our understanding of the evolution of BD magnetic activity. Figure 3
shows our current knowledge of the relationship between X-ray
emission, parameterized as $L_{X}$/$L_{bol}$ and the age of BDs. The
X-ray properties for BDs in the Orion Nebula, $\rho$ Ophiuchi and IC
348 are obtained from $Chandra$ ACIS data by Feigelson et al.\
(2002a), Imanishi, Tsujimoto \& Koyama (2001), and Preibisch \&
Zinnecker (2001), respectively. The field BD LP 944-20 detection is
from Rutledge et al.\ (2000). Individual BD ages are distributed
randomly for one order of magnitudes centered on 0.4 Myr ($\rho$ Oph,
Orion) and 1.5 Myr (IC348) based on Luhman et al. (2000). For Orion,
the upper-limits of X-ray undetected brown dwarfs were obtained as
described in Feigelson et al.\ (2002a, equations 7 and 10) assuming
negligible absorbing column and bolometric luminosities from
Hillenbrand and Carpenter (2000). The data of LP 944-20 are separately
shown for the flare, the upper limit at the quiescent (pre-flare)
phase and the time-average during full observation, while the others
are shown only for the last one (full observations). Except for LP
944-20, only two sources in $\rho$ Oph and one in IC348 have clear
flares with amplitudes less than a factor of ten, which are within the
range of the data scattering of each field. TWA 5B at $\log
L_X/L_{bol} = -3.4$ lies just below the $-$3.0 saturation limit in
main sequence clusters and is consistent with the BDs seen in younger
clusters.

Although a clear conclusion cannot be made from one object, this
result is consistent with a continued high level of BD X-ray surface
activity through $t \simeq 12$ Myr. The decay to lower levels would
then occur during the $10^7$--$10^8$ yr phase, considering the
averaged level of LP 944-20. This is consistent with the nearly
constant (or even slightly rising; see Feigelson et al. 2003) $\log
L_X/L_{bol}$ distribution from $t < 1$ Myr through $t \simeq 20$ Myr
seen in higher mass late-type stars. The high level in $\log
L_X/L_{bol}$ in younger phase might be related to the deuterium
burning, with the fact that TWA 5B lies just at the end of the burning
phase (Weintraub et al. 2000).

A somewhat different result emerges from consideration of the
relationship between X-ray plasma temperature and age in BDs (Figure
4). Here the temperatures are derived from spectral fits to $Chandra$
ACIS data by the authors listed above.  While very young BDs have
hotter temperatures above 1 keV, the intermediate-age BD TWA 5B is
significantly cooler at $kT \simeq 0.3$ keV similar to the temperature
found in the flare of the older LP 944-20 (Rutledge et al.\ 2000).
However, the low temperature in TWA 5B, combined with the lack of
variability, suggests that its X-ray emission arises more from a
`corona' than a single `flare'.

While a similar trend from hotter plasma to cooler plasma is seen in
higher-mass YSOs, the timescale for this transition is much longer
than implied for BDs. X-ray temperatures are typically $kT \simeq 6$
keV for protostars (e.g. Imanishi, Koyama, \& Tsuboi 2001) and $kT
\simeq 1-4$ keV for typical T Tauri stars (e.g. Feigelson et al.\
2002a), with highly variable flare-dominated lightcurves. This hotter
plasma is dominated by cooler coronal plasma only after $\simeq 500$
Myrs (G\"udel, Guinan \& Skinner 1997).

It is interesting that the plasma temperature in LP 944-20 flaring is
similar to that in TWA 5B quiescent. Considering that the X-ray
luminosity of the LP 944-20 flaring is at least an order of magnitudes
less than that of TWA 5B quiescent, the X-rays of TWA 5B, which appear
like coronal emission, might be explained by a superposition of a
number of such small flares. The similar possibility has been
suggested and examined for the solar corona (Parker 1988, Shimizu \&
Tsuneta 1997).

\subsection{Comments on the relationship between X-ray and H$\alpha$ emissions}

Figure 5 shows our current knowledge of the relationship between X-ray
emission, parameterized as $L_X$/$L_{bol}$ and the H$\alpha$
equivalent widths of BDs. In this plot, the ordinate indicates the
X-ray surface brightness at the BD surface, while the abscissa
represents the H$\alpha$ surface brightness, in the zero's order. Here
the H$\alpha$ properties are obtained from Jayawardhana, Mohanty \&
Basri (2002) for BDs in the $\rho$ Ophiuchi, Luhman (1999) and Herbig
(1998) for those in IC 348, and Tinney (1998) for the field BD LP
944-20. For the time-averaged X-rays of detected BDs, we obtained a
correlation of $\log(L_X/L_{bol}) = 1.5 \times \log(H\alpha) -
5.3$. Such correlation has been already known in higher mass
weak-lined T Tauri stars (for a recent example; Preibisch \& Zinnecker
2002), which is interpreted by the hypothesis that the chromosphere
(traced by the H$\alpha$ activity) is heated by a sufficient overlying
corona (X-ray activity) (Cram 1982).

Figure~5 also shows that the strongest six ($\sim$100 {\AA}) in the
H$\alpha$ equivalent width were not detected in X-rays. As with higher
mass classical T Tauri stars, these strong emission lines would be
accompanied by IR disks. The relation given in Figure~5 is thus
straightforwardly interpreted by the scenario that X-rays are mainly
due to the coronal activity while H$\alpha$ is dominated by the
chromospheric activity ($\lesssim$ a few tens {\AA}) and by the disk
activity ($\gtrsim$ a few tens {\AA}). These nondetections also
suggest that BD X-ray activity does not likely originate from
star-disk interaction but from the solar-like magnetic activity.

\section{Conclusion}

We report the detection of X-rays from an intermediate-age ($t \simeq
12$ Myr) BD at a level of $L_X \simeq 4 \times 10^{27}$ erg s$^{-1}$
(0.1--10 keV band) or $\log L_X/L_{bol} = -3.4$. The X-ray spectrum is
very soft; the dominant plasma temperature is only 0.3 keV. No
variability is seen during the 10.3 ks observation. Those
characteristics are common to those in the solar corona. Our
observation provides a link between the active state seen in younger
$t \simeq 1$ Myr stellar clusters, and a relatively inactive state,
showing continuity in the evolution of both X-ray luminosity and
plasma temperature: $\log L_X/L_{bol}$ has not yet decayed by $t
\simeq 12$ Myr while $kT$ has already substantially cooled.
The correlation between X-rays and H$\alpha$ implies the idea that the
X-rays are mainly due to the coronal activity while H$\alpha$ is
dominated by the chromospheric activity ($\lesssim$ a few tens {\AA})
and by the disk activity ($\gtrsim$ a few tens {\AA}). Finally, we
note that since TWA 5B is not far from the boundary between BDs and
the most massive planets found orbiting nearby stars, it raises the
possibility that massive planets might emit X-ray during their youth.

\acknowledgements{ We express our thanks to all the $Chandra$ team for
many efforts of the fabrication of satellite, launching, daily
operation, software developments and calibrations. We also thank Leisa
Townsley for supplying the CTI corrector of ACIS and Kensuke Imanishi
the referee Gibor Basri for useful comments on brown dwarf flares and
coronae. This research was supported by NASA contract NAS 8-38252.  }

\clearpage

\begin{figure}
 \epsscale{0.7}
\plotone{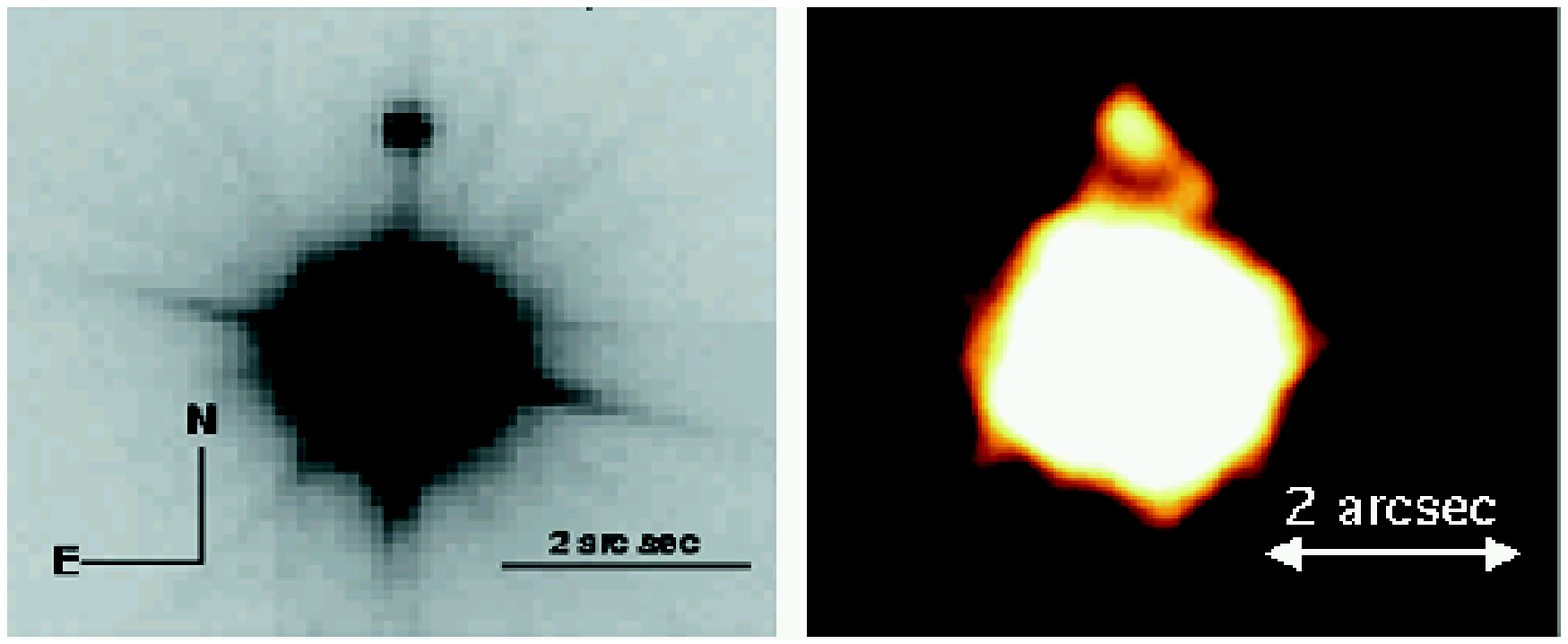}
 \epsscale{0.2}
\plotone{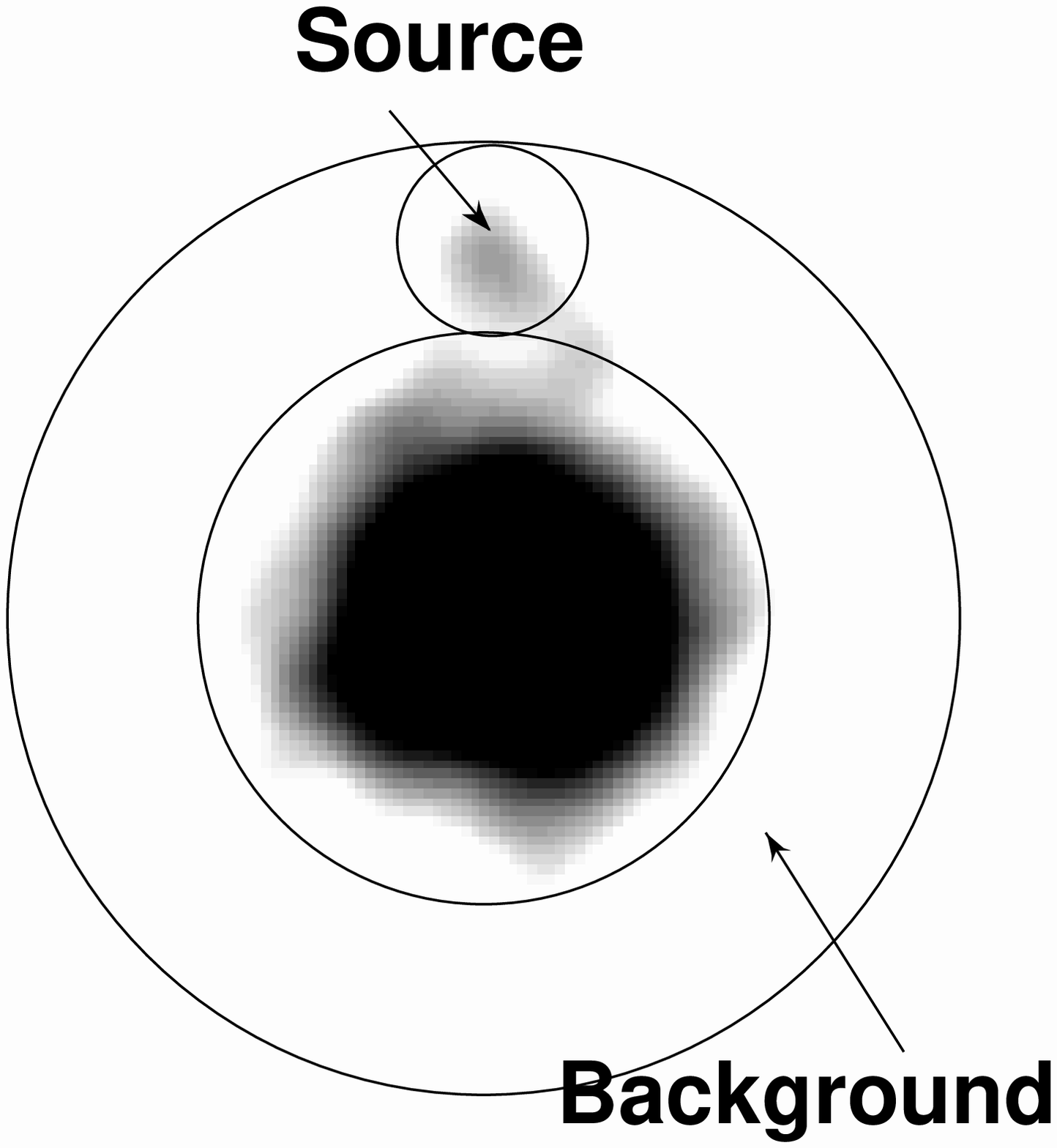}
\caption{The TWA 5 system showing the intermediate-age BD TWA 5B just
north of the primary star: (left) I-band image from Neuh\"auser et al. 
2000; (center) $Chandra$ x-ray image (0.1--1.2 keV) obtained
here; and (right) diagram of the X-ray spectral extraction regions.
The {\it Chandra} image is displayed in 0\arcsec.1 pixels after
special processing steps outlined in the text and smoothing with a
0\arcsec.5 FWHM Gaussian.}
\label{fig:lc}
\end{figure}

\clearpage 

\begin{figure}
 \epsscale{0.45}
\plotone{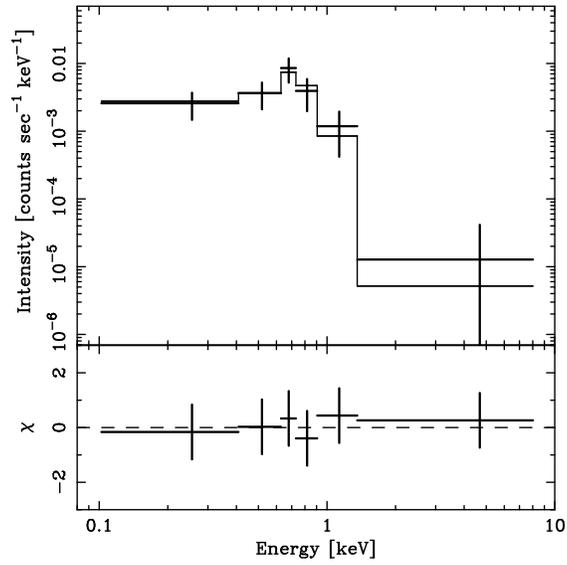}
\caption{The ACIS spectrum of TWA 5B extracted from a 0\arcsec.7 radius 
region after background subtraction (see Figure 1, right panel). 
The spectrum is fitted with a coronal plasma model, and the lower panel
shows the residuals from the best-fit model.  } \label{fig:spec}
\end{figure}

\clearpage 

\begin{figure}
\epsscale{0.45} \rotatebox{270}{ \plotone{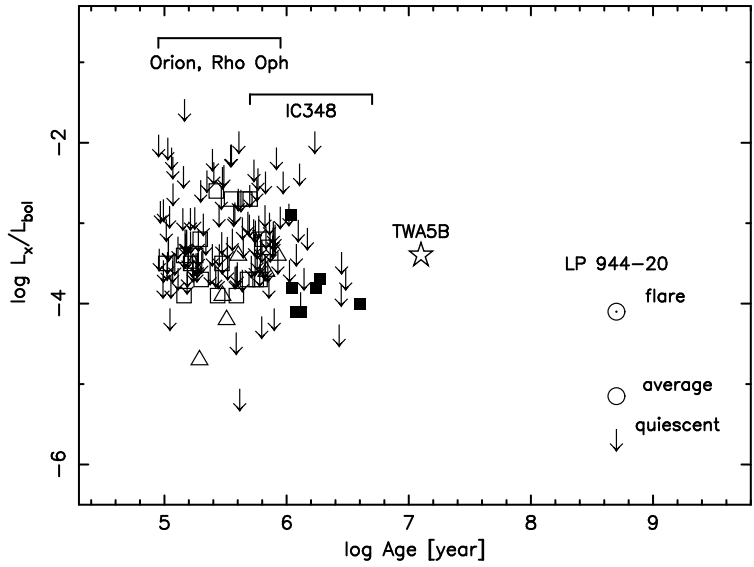}} \caption{$\log
L_X/L_{bol}$ vs. age of BDs. Symbols: Orion Nebula Cluster (open
squares), $\rho$ Ophiuchi (open triangles), and IC 348 (filled
squares). The star indicates TWA 5B. LP 944-20 is separately shown for
each period of the flare (the dotted circle), quiescent (the open
circle), and full observation (the arrow). Nondetections are indicated
with arrows.  All data are obtained from $Chandra$ ACIS observations
(see text for references).  }  \label{fig:LxLbol}
\end{figure}

\clearpage 

\begin{figure}
 \epsscale{0.45}
\rotatebox{270}{
\plotone{kT2.ps}}
  \caption{Plasma temperature vs. age of BDs. 
The symbols are the same as those in Figure 3.
}  \label{fig:kTvsAge}
\end{figure}

\clearpage 

\begin{figure}
 \epsscale{0.45}
\rotatebox{270}{
\plotone{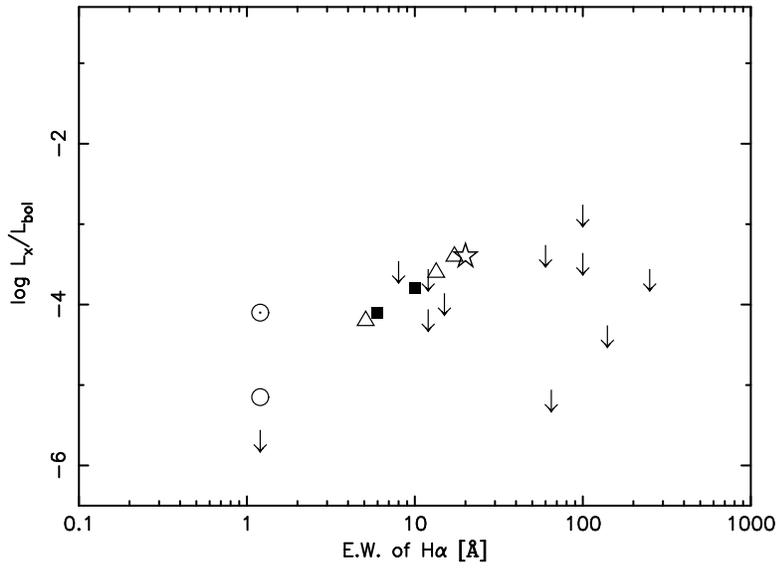}}
  \caption{$\log L_X/L_{bol}$ vs. H$\alpha$ equivalent width of BDs. 
The symbols are the same as those in Figure 3.
}  \label{fig:LxLbolvsHa}
\end{figure}

\end{document}